\documentclass[aps,prl,twocolumn]{revtex4}

\usepackage{graphicx}

\usepackage[normalem]{ulem}

\renewcommand{\b}{~}  

\pacs{03.75.Hh,05.30.Jp}

\begin{document}

\title{Ground-State Properties of a One--Dimensional System of Hard
Rods}

\author{F. Mazzanti$^1$, G. E. Astrakharchik$^2$, J. Boronat$^2$, and
J. Casulleras$^2$}

\address{ $1$ Departament de F\'{\i}sica i Enginyeria Nuclear, Comte Urgell
  187, Universitat Polit\`ecnica de Catalunya, E-08036 Barcelona, Spain}
\address{ $2$ Departament de F\'{\i}sica i Enginyeria Nuclear, Campus Nord
  B4-B5, Universitat Polit\`ecnica de Catalunya, E-08034 Barcelona, Spain}


\begin{abstract}
A quantum Monte Carlo simulation of a system of hard rods in one
dimension is presented and discussed. The  calculation is
exact since the analytical form of the wavefunction is
known, and is in excellent agreement with predictions obtained from
asymptotic expansions valid at large distances. The analysis of the
static structure factor and the pair distribution function indicates
that a solid-like and a gas-like phases exist at high and low
densities, respectively. The one-body density matrix decays following
a power-law at large distances and produces a divergence in the low
density momentum distribution at $k=0$ which can be identified as a
quasi-condensate.

\end{abstract}

\pacs{03.75.Hh, 67.40.Db}

\narrowtext

\maketitle


Correlated (quasi)-one-dimensional (1D) systems of bosons and fermions
have received great attention in the last years due to recent and
important experimental progress~\cite{Paredes2004, Bloch2005,
Moritz2003, Richard2003}.  The role of quantum fluctuations is
enhanced in reduced dimensionalities, producing new and intriguing
features different or not present in 3D systems. A well known but
striking difference is the nonexistence of a true Bose condensate in
1D homogeneous systems at any temperature~\cite{Hohenberg1967} (not
even at $T=0$), although it can be realized in trapped systems where
the confining potential modifies the atomic density of
states~\cite{Gorlitz2001}.

Experimentally, 1D systems
can be realized by
confining the radial motion of a 3D trapped cloud of cold bosons to
zero point oscillations. This is done by acting on the system with two
orthogonal standing waves that create an optical lattice containing an
array of 1D quantum gases in the axial direction. The ensemble
generated in this way allows for a statistical treatment of the
relevant quantities being measured. A few years ago,
Olshanii~\cite{Olshanii1998} showed that in these experiments the
scattering length $a_{1D}$ of the resulting systems experience a
confined induced resonance\b according to the expression
\begin{equation}
a_{1D} = {a_\perp^2 \over a_{3D}}\left( 1 - C {a_{3D} \over a_\perp}
\right) \ ,
\label{a1d}
\end{equation}
where $a_{3D}$ is the 3D scattering length of the interatomic
potential, $a_\perp=\sqrt{\hbar/m\omega_\perp}$ is the oscillator
length of the transverse confinement, and
$C=\zeta(1/2)/\sqrt{2}=1.0326$ with $\zeta(\cdot)$ the Riemann zeta
function. In actual experiments, $a_{3D}$ can be tuned to essentially
any value in the range $(-\infty,+\infty)$ by exploiting a Feshbach
resonance, and thus $a_{1D}$ can be made to vary accordingly, as seen
from Eq.~(\ref{a1d}). In a pseudopotential description, where $a_{1D}$
is directly related to the coupling constant $g_{1D}$ of the contact
interaction $U(z)=g_{1D}\delta(z)$ through the relation
$g_{1D}=-2\hbar^2 / ma_{1D}$~\cite{Olshanii1998}, different regimes
can be realized when $a_{1D}$ or the density $n$ are changed.  These
regimes can be classified in terms of the ratio of the interaction
energy per particle in a mean-field approximation,
$g_{1D}n=-2\hbar^2n/ma_{1D}$, to the characteristic kinetic energy per
particle $\hbar^2n^2/2m$. When $n a_{1D}$ is large $g_{1D}$ is small,
the effect of correlations is weak and the system enters a mean field
regime.  As $g_{1D}$ increases, $n a_{1D}$ decreases and potential
effects are more relevant.  In the $n a_{1D}\to 0$ limit, $g_{1D}$
goes to infinity and the system becomes a Tonks-Girardeau gas of
impenetrable bosons~\cite{Girardeau1960}. In this regime, correlations
are so strong that the ground-state wave function acquires a fermionic
behavior and it vanishes when two or more particles
meet~\cite{Girardeau1960,Lieb1963}.  More recently, a new state called
super-Tonks-Girardeau, corresponding to $g_{1D}\to-\infty$, has been
identified and shown to exhibit even stronger correlation
effects~\cite{Gregory2005}. In this regime the system behaves as a gas
of hard rods of length $a=a_{1D}$ for particle densities $n a_{1D}\leq
0.1$.  At higher densities, the interatomic potential of a system of
hard rods is no longer {\em weak} for a pseudopotential picture to be
realistic. Nevertheless, at high densities the hard rods model can be
used to understand static and dynamic properties of strongly
correlated 1D systems with higly repulsive interactions at short
distances, like He or other gases adsorbed in Carbon
nanotubes~\cite{Pearce2005, Mercedes2001}.

Hard rods are the 1D counterpart of hard spheres in
3D~\cite{Giorgini1999,Mazzanti2003}. The interatomic rod potential reads
$V_{HR}(z)=+\infty$ for $|z|<a$ and 0 otherwise.
The associated $N$--particle Hamiltonian becomes
\begin{equation}
H = -{\hbar^2 \over 2m} \sum_{j=1}^N {\partial^2 \over \partial z_j^2}
+ \sum_{i<j} V_{HR}(z_{ij}) \ .
\label{Hamiltonian}
\end{equation}
whith the {\em exact} ground-state wavefunction~\cite{Nagamiya1940}
\begin{equation}
\Psi_0(z_1, z_2,..., z_N) = {1\over \sqrt{N!}} \left|
{\rm det}\left( {1\over\sqrt{L'}} \exp( i p_k' x_k ) \right)\right| \ .
\label{wavefunction}
\end{equation}
In this expression, $L'=L-a N$ is the {\em unexcluded} volume,
$\{p_k'=2\pi n_k/L'\}$ are a set of quantum numbers with
$n_k\in[-N,+N]$,
and $x_k=z_k - (k-1) a$ are the so--called rod coordinates
corresponding to a given ordering of the true particle coordinates
$z_1< z_2-a < z_3 - 2a < \cdots < z_N - a(N-1)$. As in the 3D case of
hard spheres, the scattering length of the hard rod potential equals
the size of the rod, $a_{1D}=a$.

Despite the fact that the analytical form of the groundstate
wavefunction is known, limited progress has been achieved in the
description of this system~\cite{Rubin1955,Krotscheck1999}. In this
work, we analyze and discuss the static properties of a gas of hard
rods of length $a$ at $T=0$ as a function of the density by means of
Monte Carlo simulations.  We sample the wavefunction
(\ref{wavefunction}) using the Metropolis algorithm \b and impose
periodic boundary conditions for a number of particles in the range
$N\leq 300$.  Notice that since this wavefunction is the exact
solution to the $N$--body problem corresponding to the Hamiltonian in
Eq.~(\ref{Hamiltonian}), the results of the simulations are
statistically exact.
As a check to the calculation we reproduce numerically with zero
variance the equation of state of the system, whose analytical
expression reads
\begin{equation}
{E_{HR}\over N} = {\pi^2 \hbar^2 n^2 \over 6m} {1 \over (1 - n a)^2} \ .
\label{EnergyHR}
\end{equation}

We first analyze \b the static structure factor $S(k)=\langle \Psi_0
\mid\rho^\dagger_k \rho_k \mid \Psi_0\rangle/N$, with
$\rho_k=\sum_{j=1}^N e^{i k z_j}$ the density fluctuation operator.
Even though the ground-state wave function is known, no simple
analytical expression can be easily derived for $S(k)$, although three
notorious properties can be inferred. On one hand, the system is a
realization of a Luttinger liquid with low $k$ excitations dominated
by phonons~\cite{Haldane1981}, and therefore $S(k\to 0)= \hbar |k| / 2
m c$. The speed of sound $c$ can be obtained from the equation of
state~(\ref{EnergyHR}) and leads to
\begin{equation}
S(k\to 0) \approx {(1-a n)^2 \over 2\pi n}\,|k| \ .
\label{skto0}
\end{equation}
On the other hand and for a given particle ordering, the density
fluctuation operator becomes $\rho_k = \sum_{j=1}^N e^{i k
x_j}$ when $k$ is a multiple of $2\pi/a$. In this case, the $a$
factors in the change to rod coordinates $\{z_k\}\to\{x_k\}$ have no
influence in $S(k)$, which equals the corresponding value of the
static structure factor of the 1D free Fermi gas (FFG) at the
rod density $n'=n/(1-a n)$
\begin{equation}
S_{FFG}(k) = \left\{
\begin{array}{ll}
{1 - a n \over 2\pi n}\,|k| & \mbox{for $|k|\leq {2\pi n \over 1 - a
n}$} \\ 1 & \mbox{ otherwise \ .}
\end{array}
\right.
\label{Sk-1DFFG}
\end{equation}
Introducing explicitly the change to rod coordinates, $S(k)$ can
be written in the exact form
\begin{equation}
S(k)\!\!=\!\! 1 +2(N\!-\!1)!\!\!\sum_{i=1}^{N-1}\!\sum_{j=1}^{N-i}
\!\!\int_{\Omega_N}
\!\!\!\!\!\!
dx^N\!\!\cos\left[ k(x_{i+j,i}\!\!+\!\!ja) \right]\! \Psi^2_0
\label{Skint}
\end{equation}
where $x_{i,j}=x_i-x_j$, $\Omega_N$ denotes the integration region
$0\leq x_1<x_2<\cdots<x_N\leq L'$ and $\Psi_0$ is the Slater
determinant of Eq.~(\ref{wavefunction}). Due to strong correlations
the most probable configurations are those where all particles are
equally spaced at a distance $\Delta x=L'/N$. 
The contribution to $S(k)$ becomes maximal when all the cosine terms
in Eq.~(\ref{Skint}) equal $1$ for these configurations, which happens
at the discrete values $k_j=2\pi n j$.  In summary, one expects an
$S(k)$ growing linearly at low $k$, presenting an infinite number of
equally spaced maxima, and approaching the asymptotic value of 1 when
$k\to\infty$.
\begin{figure}[t!]
\begin{center}
\includegraphics*[width=0.41\textwidth]{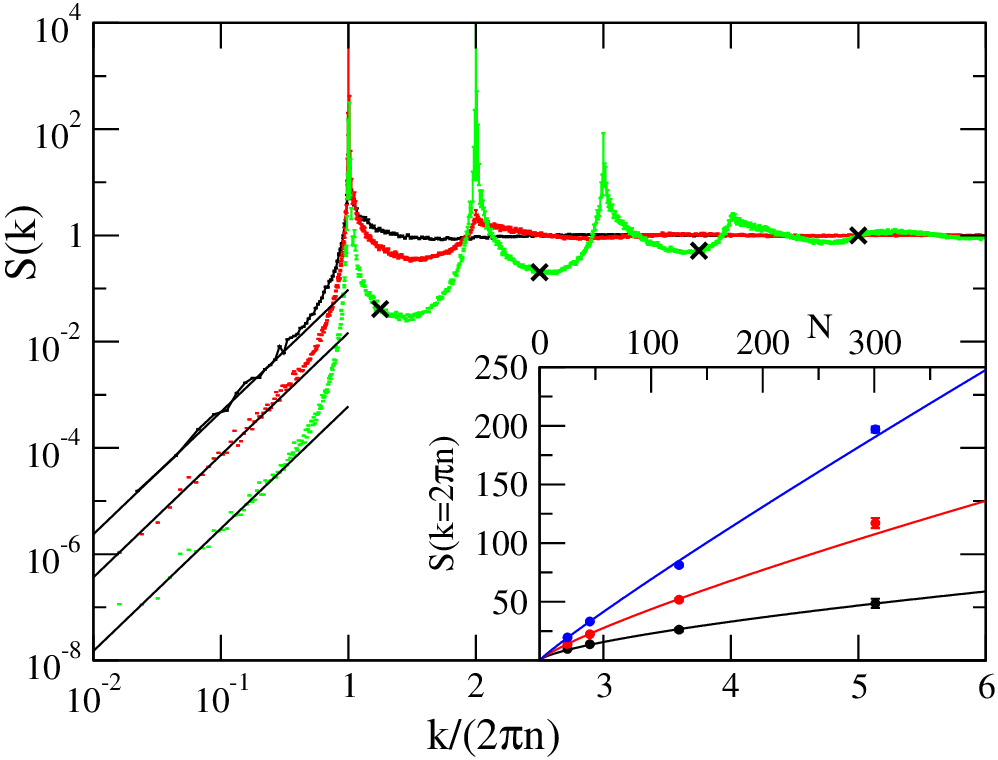}
\end{center}
\caption{ (Color online) $S(k)$ at particle densities $n a=0.4$, $0.6$
and $0.8$ (upper, middle and lower cures at low $k$). Notice the
logarithmic scale up to $k/2\pi n=1$. Solid line: phononic behavior of
Eq.~(\ref{skto0}). The crosses correspond to the exact values from
Eq.~(\ref{Sk-1DFFG}), for $na=0.8$ at $k_i=2\pi i /a,
i=1,2,3,4$. Inset: height of the first peak as a function of the total
number of particles (symbols) for densities $na=0.6, 0.7$ and $0.8$
(lower, middle and upper curves, respectively), best fit with the law
given by Eq.~\ref{SkPeak} and $m=1$ (lines).}
\label{fig-sk}
\end{figure}

Results for $S(k)$ at three different densities are shown in
Fig.~\ref{fig-sk}.  A logarithmic scale up to $k/2\pi n=1$ has been
used to emphasize the $k\to 0$ linear behavior of $S(k)$ given in
Eq.~(\ref{skto0}).  The upper, middle and lower curves in that region
correspond to $n a=0.4, 0.6$ and $0.8$, respectively. At lower
densities $n a\leq 0.1$, $S(k)$ is much smoother and approaches the
hard point limiting case of Eq.~(\ref{Sk-1DFFG}).  The crosses in the
plot correspond to the exact values obtained from this equation at
$k_i=2\pi i/a$. As it can be seen from the figure, the
peaks at $k_j=2\pi n j$ are enhanced at higher densities while at
intermediate values the strength is depressed.  Our simulations
indicate that the height of the peaks increases with the particle
number at large densities. This dependence can be understood by
looking at the asymptotic expansion of the pair distribution function,
the Fourier transform of $S(k)$,
which admits, for a Luttinger liquid and according to
Haldane~\cite{Haldane1981}, the following asymptotic expansion valid
when $|z|\gg n^{-1}$
\begin{equation}
g(z) = 1 - {\eta \over (2\pi n z)^2} + \sum_{m=1}^\infty
A_m {\cos(2\pi n\, m\, z) \over (n |z|)^{m^2 \eta}} \ ,
\label{Haldane-gr}
\end{equation}
where $\eta=2K$ and $K=\pi\hbar n/mc$ is the Luttinger parameter, while
the coefficients $A_m$ depend both on the density and the system under
study. Notice that the $z^{-2}$ term coming from density-density
fluctuations determines the low $k$ behavior of $S(k)$ reported in
Eq.~(\ref{Skint}). Furthermore, $\eta=2(1-na)^2$ for a system of hard
rods while $\eta=2$ for the 1D free Fermi gas.  According to this
expression, the height of the $m$-th peak follows 
a power-law of the form $|z|^{1-m^2\eta}$. The inset in
Fig.~\ref{fig-sk} shows the height of the
first peak as a function of the number of particles in the simulation
compared with the corresponding curves
\begin{eqnarray}
S(k=2m\pi n) = A_m N^{1-2m^2(1-na)^2},
\label{SkPeak}
\end{eqnarray}
with $m=1$ corresponding to the first peak.  A fit to the Monte Carlo
data shows that these curves are compatible with the choice $A_1=1$ at
high densities.  Furthermore, the law (\ref{SkPeak}) predicts that
only a finite number of macroscopic peaks located at $k_m=2\pi n m$
and satisfying the inequality $1-2m^2(1-na)^2>0$ exist.  In the case
$na=0.8$ this implies that $m<3.5$, and we find only three peaks whose
height grows with the number of particles.  The linear behavior
$S(k)\propto N$ at the peak, characteristic of 3D crystals, is
recovered asymptotically as $na\to 1$.  All these facts suggest
that a packing order, resulting from the combined effect of particle
correlations and the reduced dimensionality, shows up at high
densities, thus manifesting the existence of a {\em quasi--solid}
phase.  At low density these effects, although present, are much less
evident, as the peaks are washed out at $na \ll 1$ and $S(k)$
approaches the simple structure corresponding to a 1D free Fermi gas
with the density $n'=n/(1-an)$ reported in Eq.~(\ref{Sk-1DFFG}). In
this sense, the system of hard rods clearly presents different regimes
and behaves as a {\em quasi--solid} at high densities.

\begin{figure}[t!]
\begin{center}
\includegraphics*[width=0.41\textwidth]{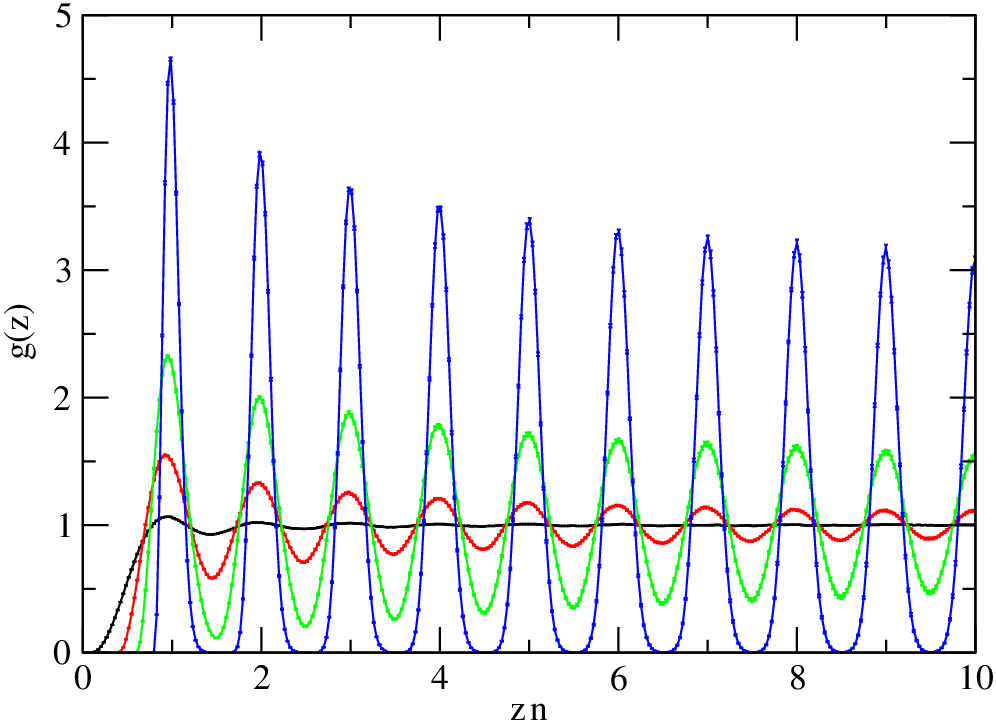}
\end{center}
\caption{(Color online) Two-body radial distribution function at different
  densities. From top to bottom: $na=0.8$, $0.6$, $0.4$ and $0.1$,
  respectively.}
\label{fig-gz}
\end{figure}

The pair distribution function $g(z)$ is depicted in
Fig.~\ref{fig-gz} for several densities.
Being related to $S(k)$ by a Fourier
transformation, it reproduces the same packing structure and
particularly presents a series of peaks located at multiples of $1/n$,
coming from the $m=1$ term in Eq.~(\ref{Haldane-gr}).  As in the case
of $S(k)$, the strength of the peaks increase with the density, while
in all situations $g(z)=0$ inside the core of the potential 
$|z|<a$.  At the lowest densities, $g(z)$ approaches the hard point
limit corresponding to the 1D free Fermi gas
\begin{equation}
g_{HP}(z) = 1 - {\sin^2(\pi n z) \over N^2\sin^2(\pi n z/N)}
\label{gz1DFFG}
\end{equation}
with $n'$ replaced by $n$ since the distinction between them is not
important at low densities.
This function
presents a periodic structure with $N/2$ equally spaced peaks in the
range $[0,L/2]$.
In the thermodynamic limit, $g_{HP}(z)$ admits an
expression of the form~(\ref{Haldane-gr}) with $\eta=2$ and
$A_m=(2\pi^2)^{-1}\delta_{m,1}$, with a single frequency contributing
to the oscillations.  In the case of hard rods, the peaks remain at
the same location and are enhanced as the density is increased, which
indicates that the number of peaks extends to infinity in the
thermodynamic limit.

\begin{figure}[b!]
\b
\begin{center}
\includegraphics*[width=0.41\textwidth]{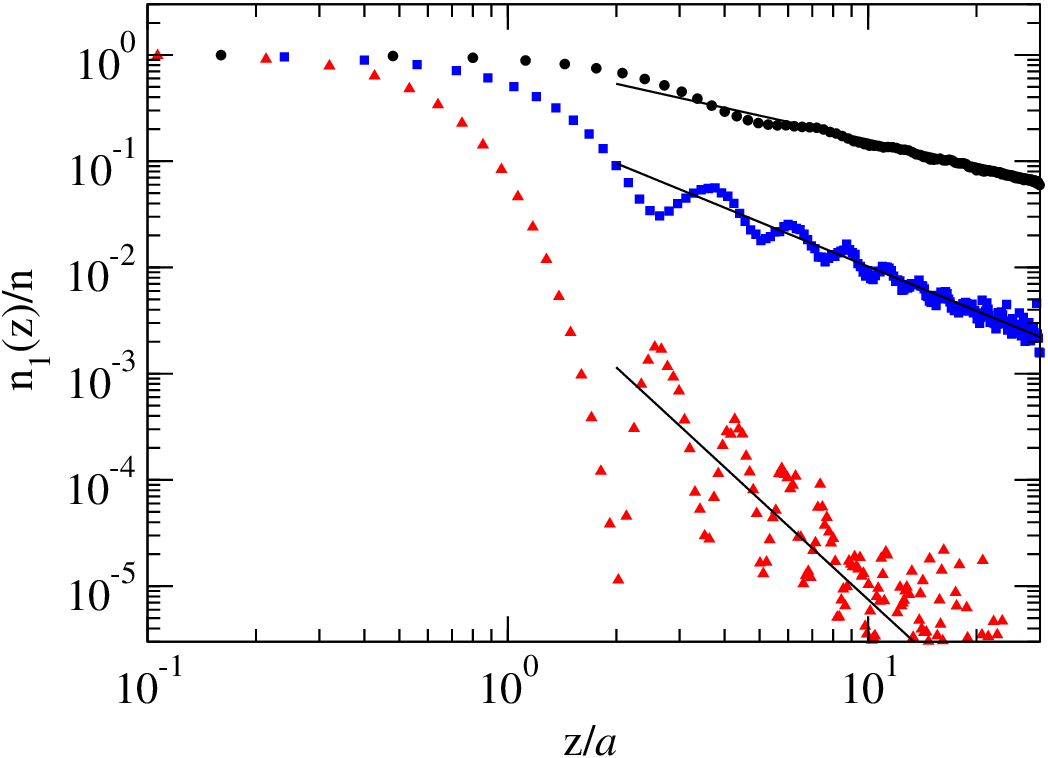}
\end{center}
\caption{(Color online) One--body density matrix at the densities
  $na=0.2, 0.4$ and $0.6$ (upper, middle and lower curves,
  respectively). The solid lines represent the asymptotic behavior at
  long distances.}
\label{fig-rho1}
\end{figure}

The next quantity analyzed is the off-diagonal one--body density matrix.
From translational invariance arguments and the normalization of the
wave function it follows that $n_1(0)=n$.  In systems of higher
dimensionality, the presence of a Bose condensate with density $n_0$
induces non--diagonal long range order that is manifested in a finite
asymptotic value $n_1(|z|\to\infty) = n_0>0$.  This is not the case in
homogeneous 1D systems, where a true BEC is suppressed and thus
$n_1(|z|\to\infty)\to 0$.  Figure~(\ref{fig-rho1}) shows the one--body
density matrix in logarithmic scale and for the three densities
$na=0.2$, $0.4$ and $0.6$. Clearly, $n_1(z)$ shows an oscillating
structure that expresses the presence of an excluded length
corresponding to the rod size.

Uniform Bose Luttinger liquids admit an asymptotic expansion valid at
large distances of the form~\cite{Haldane1981}
\begin{equation}
{n_1(z)\over n} = {1\over(n|z|)^{1/\eta}}
\sum_{m=0}^\infty B_m { \cos(2\pi n\,m\,z) \over (n|z|)^{m^2\eta}} \ ,
\label{Haldane-rho1}
\end{equation}
and thus the long range behavior of $n_1(z)$ depends on the value of
the Luttinger parameter $\eta$. For hard rods this means that
$n_1(|z|\gg 1)$ decays following a $|z|^{-1/2(1-an)^2}$ power law.  A
fit to the tail of the data of this analytical form is also shown in
Fig.~(\ref{fig-rho1}) for each density.

\begin{figure}[b!]
\begin{center}
\includegraphics*[width=0.41\textwidth]{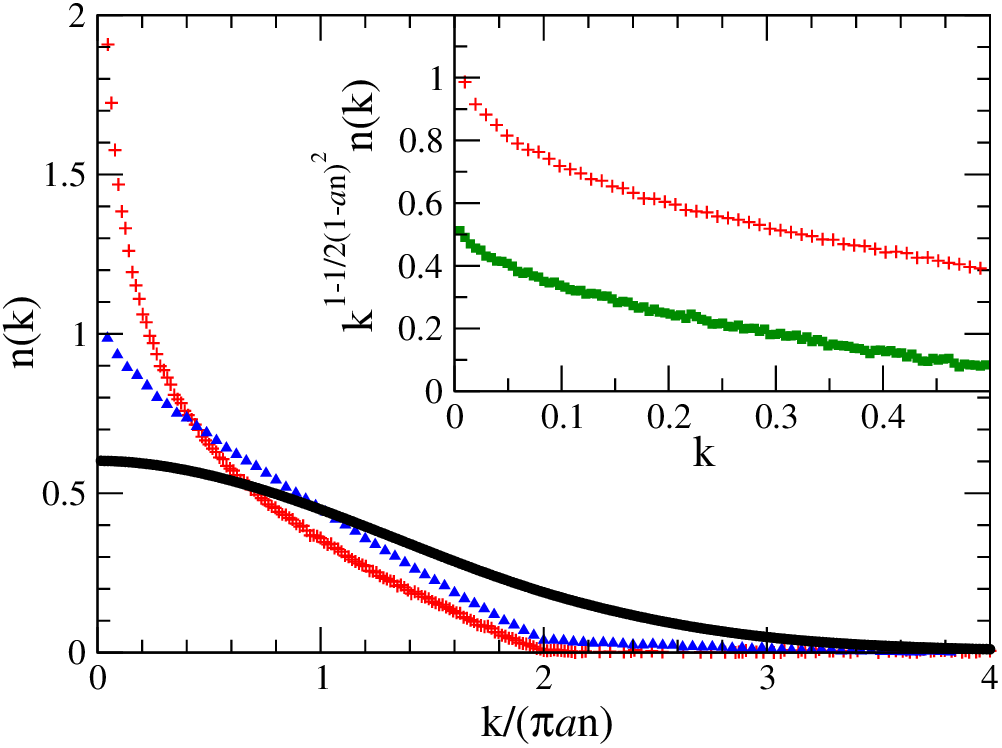}
\end{center}
\caption{(Color online) Momentum distribution $n(k)$ of the hard rods
  system at the densities $na=0.2, 0.4$ and $0.6$ (upper, middle and
  lower curves).  The inset shows the product $k^{1-1/\eta} n(k)$ at
  $na=0.1$ (squares) and $na=0.2$ (stars).}
\label{fig-nk}
\end{figure}

The momentum distribution $n(k)$, which describes the occupation of
each single--particle state of momentum $k$, is the Fourier transform
of the one--body density matrix.  Fig.~\ref{fig-nk} displays $n(k)$
for three different densities. The large $z$ power law decay of the
one--body density matrix makes the low $k$ behavior of $n(k)$ depend
on $\eta$.  For hard rods $\eta=2(1-na)^2$ and that dependence defines
a critical density $n_ca=1-1/\sqrt{2}\approx 0.29$ separating two
different regimes. At lower densities the momentum distribution
presents an infrared divergence of the form $k^{1-1/2(1-na)^2}$ and a
discontinuity in the first derivative at $k=2\pi n$, as can be checked
by direct inspection of the Fourier transform of the $m=0$ and $m=1$
terms in Eq.~(\ref{Haldane-rho1}). Both features disappear at higher
densities, although a change in the slope of $n(k)$ is still
noticeable at intermediate values of $na$.
The inset in Fig.~\ref{fig-nk} shows the product
$k^{1-1/2(1-na)^2}n(k)$ for two densities.
The divergence of $n(k=0)$ can be interpreted
as the manifestation of a Bose--Einstein quasi--condensate, while the
kink at $k=2\pi n$ is a reminiscence of the underlaying fermionic
nature of the wave function, as for 1D Fermions $k_F=\pi n$.

We end up this discussion by noticing that the arguments leading to
the exact values of $S(k)$ at $k_j=2\pi j/a$ can be extended to
predict the behavior of the $T=0$ dynamic structure function
$S(k,\omega)$ at these same momenta, which equals that of the 1D
free Fermi gas at the equivalent density $n'=n/(1-an)$
\begin{equation}
 S\left( k_j={ 2\pi j\over a},\omega \right) =
\left\{
\begin{array}{lcl}
{1 - an \over 2\pi n}\left( {m\over
 k_j} \right) & ; & \omega\in(\omega_0, \omega_1) \\
 0 & & \mbox{otherwise}
\end{array}
\right.
\end{equation}
with $\omega_0=\mid k_j^2/2m - \pi k_j n/(1-an)\mid$ and
$\omega_1=k_j^2/2m + \pi k_j n/(1-an)$. In this way, $S(k=k_j,\omega)$
becomes a constant independent of $\omega$ in the range $(\omega_0,
\omega_1)$, and leads to the finite values of $S(k)$ reported in
Eq.~(\ref{Sk-1DFFG}) once integrated. At different momenta $k\neq
k_j$, $S(k,\omega)$ is expected to present at least one peak that
increases when $k$ approaches the values $k_i=2\pi n i$ where $S(k)$
diverges in the thermodynamic limit.

In summary, we have carried out a complete study of the most relevant
one- and two- body correlation functions for a system of hard-rods at
$T=0$.  We find two distinct regimes where the system behaves as a gas
(low density) and as a quasi-solid (large density), without any
signature of a phase transition in the energy. The quasi-solid
regime is characterized by the presence of macroscopic peaks in the
static structure factor.  The one--body density matrix at large
distances decays following a power law that leads to a divergence of
the low density momentum distribution at $k=0$. This divergence can be
understood as the manifestation of a Bose--Einstein quasi-condensate.
Finally, exact values for the static structure factor and dynamic
structure function at the momenta $k_j=2\pi j/a$ have also been
reported.  
Our results allow for a much better understanding of the fundamental
hard rod model. 
We hope our work can stimulate further
experimental work both in dilute vapors and in condensed phases in 1D
systems. 



\begin{acknowledgments}
This work has been partially supported by Grants No. FIS2005-04181 and
FIS2005-03142 from DGI (Spain), and Grant No. 2005SGR-00779 from the
Generalitat de Catalunya.  G.E.A. acknowledges useful discussions with
Dr.~D.Gangardt and support from MEC (Spain).

\end{acknowledgments}



\begin{references}

\bibitem{Paredes2004} B. Paredes {\it et al.}, Nature {\bf 429},
277 (2004).

\bibitem{Bloch2005} I. Bloch, Nature Phys. {\bf 1}, 23 (2005).

\bibitem{Moritz2003} H. Moritz {\em et al},
Phys. Rev. Lett. {\bf 91}, 250402 (2003).

\bibitem{Richard2003} S. Richard {\em et al.}, Phys. Rev. Lett. {\bf
91}, 010405 (2003).

\bibitem{Hohenberg1967} P. C. Hohenberg, Phys. Rev. {\bf 158}, 383 (1967).

\bibitem{Gorlitz2001} A. G\"orlitz, {\em et al.},
Phys. Rev. Lett. {\bf 87}, 130402 (2001).

\bibitem{Olshanii1998} M. Olshanii, Phys. Rev. Lett. {\bf 81}, 938 (1998).

\bibitem{Girardeau1960} M. Girardeau, J.Math.Phys. {\bf 1}, 516 (1960).

\bibitem{Lieb1963} E. H. Lieb and W. Liniger, Phys. Rev. {\bf 130},
1605 (1963).

\bibitem{Gregory2005} G. E. Astrakharchik {\em et al},
Phys. Rev. Lett. {\bf 95}, 190407 (2005).

\bibitem{Pearce2005} J. V. Pearce {\em et al},
Phys. Rev. Lett. {\bf 95}, 185302 (2005).

\bibitem{Mercedes2001} M. Mercedes {\em et al.}, Rev. Mod. Phys. {\bf
73}, 857 (2001).

\bibitem{Giorgini1999} S. Giorgini, J. Boronat, and J. Casulleras,
Phys. Rev. A{\bf 60}, 5129 (1999).

\bibitem{Mazzanti2003} F. Mazzanti, A. Polls, and A. Fabrocini,
  Phys. Rev. A{\bf 67}, 063615 (2003)

\bibitem{Nagamiya1940} T. Nagamiya, Proc. Phys. Math. Soc. Jpn. {\bf
22}, 705 (1940).

\bibitem{Rubin1955} R. J. Rubin, J. Chem. Phys. {\bf 23}, 1183 (1955).

\bibitem{Krotscheck1999} E. Krotscheck, M. D. Miller and J. Wojdylo,
  Phys. Rev. B{\bf 60}, 13028 (1999).

\bibitem{Haldane1981} F. D. M. Haldane, Phys. Rev. Lett. {\bf 47}, 1840 (1981).

\end{references}
 \end{document}